\documentclass{article}
\begin{document}
\baselineskip7mm
\title{Transient chaos in scalar field cosmology on a brane}
\author {A.Toporensky}
\date{}
\maketitle
\hspace{8mm} {\em Sternberg Astronomical Institute,
Universitetsky prospekt, 13, Moscow 119992, Russia}

\begin{abstract}
We study cosmological dynamics of a {\it flat} Randall-Sundrum brane
with a scalar field and a negative "dark radiation" term. It is shown
that in some situations the "dark radiation" can mimic spatial curvature
and cause a chaotic behavior which is similar to chaotic dynamics
in {\it closed} Universe with a scalar field.
\end{abstract}

%\section{Introduction}
The phenomenon of  transient chaos in homogeneous
cosmological models had been described by D. Page \cite{Page}
(he studied a closed isotropic Universe with a massive scalar field) even
earlier than this concept was formulated and investigated
systematically (see, for example, \cite{tran,Gaspard}). The key
feature of this type of chaos is that the dymamical system (in comparison
with the well-known case of strange attractors) 
has a regular regime as its future attractor while particular trajectories
can experience a chaotic behavior before reaching this stable regime.
The final outcome can be also represented by some another
situation which can be treated as a "final state" (as in the case of a
cosmological
singularity where the entire dynamics brakes down).

In the described dynamics of thr Universe a cosmological singularity is the ultimate fate
of any (except for a set of zero measure) trajectory, though 
Universe can
go through an arbitrary number of "bounces" (i.e. transitions from
contraction
to expansion) before  final contraction stage ends in a singularity.
The set of initial conditions leading to bounces has a rather regular
structure \cite{Star}, 
which allows calculation of topological entropy \cite{Cornish,four} (we
should, however,
mention that for sufficiently shallow scalar field potentials this
simple structure
of the chaos becomes more complicated \cite{shallow}).
This type of dynamics is different from the 
Mixmaster chaos where shear variables
experience chaotic
oscillations while volume of the Universe decreases monotonically. Similar
picture exists for 
chaos in two-field system, described in Ref. \cite{Easther} and for
non-abelian field
dynamics \cite{Levin,Galtsov} -- both these cases do not require volume
oscillations, which
are crucial for describing type of transient chaos.

It also differs from the chaos in a closed Universe with a conformal
massive scalar field \cite{conformal,conformal2}. The main feature of the latter
system is that
the dynamics can be prolonged through a cosmological singularity to
the range
of negative scale factors. As a result, we have chaotic oscillations of
scale
factor (it changes its sign twice during one oscillation)
without any future stable regime, and this chaos can not be treated as
"transient". Moreover, as only the
part of a trajectory before the first singularity have a physical
significance, there are claims that such
physical system (in contrast to its mathematical model) has no chaotic
properties \cite{eti,Motter}.

The equations of motion for a closed isotropic Universe with a minimally
coupled
scalar field have the form (see, for example, \cite{we})
\begin{equation}
\frac{m_P^2}{16 \pi}\left(\ddot a+\frac{\dot a^2}{2a}+\frac{1}{2a}\right)+
\frac{a \dot \varphi^2}{8}-\frac{aV(\varphi)}{4}=0
\end{equation}
\begin{equation}
\ddot \varphi+\frac{3\dot \varphi \dot a}{a}+V'(\varphi)=0
\end{equation}
with the first integral
\begin{equation}
-\frac{3m_P^2}{8 \pi}\frac{\dot a^2}{a^2} + \frac{\dot \varphi^2}{2}=
\frac{3 m_P^2}{8 \pi}\frac{1}{a^2}-V(\varphi).
\end{equation}

Here $m_P$ is the Planck mass, $a$ is the scale factor, $\varphi$ is the
scalar field
with a potential $V(\varphi)$.

A peculiar form of the first integral ($\dot a^2$ and $\dot \varphi^2$
enters in the LHS
of (3) with opposite signs) leads to some dynamical features, which
distinguish the system
(1)-(3) from other abovementioned cosmological chaotic dynamical
systems. First of all,
there are no forbidden regions in the configuration space $(a,
\varphi)$. Instead, it divided
into zone where RHS of (3) is positive (and possible extrema of the
scale factor are located),
and zone where RHS of (3) is negative (zone of possible extrema of the
scalar field). These two
zones are separated by the curve \cite{KKT}
\begin{equation}
a^2 =\frac{3}{8\pi}\frac{m_P^2}{V(\varphi)}
\end{equation}
which can be treated as a set of possible zero-velocity ($\dot a = \dot
\varphi=0$) points.
Numerical studies show that trajectories with these points play an
important role in the 
described chaotic structure. In particular, all primary (i.e.
having one bounce per period)
trajectories have zero-velocity points as the points of bounce (see
numerical examples in \cite{Cornish}).

Numerical integrations show also that there are regions on the curve (4)
which can not contain
points of bounce. If a trajectory, starting from the curve (4) is
directed inside the zone of possible
extrema of the scale factor, it rapidly goes through a point of maximal
expansion and evolves
further towards a singularity. The condition for a trajectory to be
directed into the
opposite zone (the zone of
possible extrema of the scalar field) can be written as
\begin{equation}
\ddot \varphi/\ddot a > d\varphi(a)/da
\end{equation}
where the function $\varphi(a)$ in the RHS is the equation of the curve
(4).
 
The case of equality in (5) corresponds to a trajectory, tangent to the
curve (4).
This situation was first described in \cite{Page}, and we call such point
as a Page point. For the system (1)-(3) the equation for the Page point
is \cite{we}
\begin{equation}
V(\varphi_{page})=\sqrt{\frac{3 m_P^2}{16 \pi}}V'(\varphi_{page})
\end{equation}

For power-law scalar field potentials the
condition (5) is satisfied if $\varphi > \varphi_{page}$,
and the corresponding
part of the curve (3) contains zero-velocity bounce points of periodical
trajectories. For exponential
potentials the condition (5) can be violated for all points on the curve
(3), and the whole chaotic
structure disappears \cite{we, recent}.

In all our previous studies we were interested only in steepness of the
scalar field potential $V(\varphi)$
for large $\varphi$ and its influence on the possibility of bounces. On
the other hand, any positive
potential with $V(0)=0$ in a close Universe leads to a recollaps
ultimately, while open and flat
Universe will expand forever. This is the reason why the transient chaos
exists only for closed Universe in the standard cosmology.
 However, violation of positive energy
condition can change this situation \cite{Wheeler}. There are several
possible sources of an effective negative energy in modern cosmological
scenarios. The influence of a phantom field \cite{Caldwell} on chaotic
properties of the Universe have been studied in \cite{Szydlo, Szydlo2}.
Another  possible source is so called
"dark radiation" which
appears in braneworld scenarios. The sign of dark radiation is not fixed
in the theory, and
in the case of a negative sign the dark radiation can cause the
recollaps of a {\it flat} brane
Universe. The goal of the present communication is to study the
possibility of a transient
chaos in a flat brane Universe, where recollaps is achieved solely by a
negative dark radiation.

%\section{Main text}

From now on we study a flat RS brane with a scalar field. The equations
of motions are \cite{Langlois,Langlois2}
\begin{equation}
\frac{\ddot a}{a}+\frac{\dot
a^2}{a^2}=-\frac{k^4}{36}\rho_b(\rho_b+p_b)-\frac{k^2}{6}\Lambda
\end{equation}
\begin{equation}
\frac{\dot
a^2}{a^2}=\frac{k^2}{6}\Lambda+\frac{k^4}{36}\rho_b^2+\frac{C}{a^4}
\end{equation}
Here $k^2=8\pi/M_{(5)}^3$, where $M_{(5)}$ is a fundamental
5-dimensional Planck mass, $C$ is the "dark radiation".
The matter density on a brane is
$$
\rho_b=\dot\varphi^2/2 + V(\varphi) +\lambda,
$$
where $\lambda$ is the brane tension,
the effective pressure is
$$
p_b=\dot\varphi^2/2-V(\varphi),
$$
the Klein-Gordon equation for a scalar field (2) remains unchanged.

In the eq.(7)-(8) $\Lambda$ is the cosmological constant in a bulk, and
we assume
that $\Lambda=-(k^2/6)\lambda$ (the Randall-Sundrum constraint) in order to
get the effective cosmological constant on a brane vanishing.

The cosmological dynamics on the brane depends on the ratio
$ \rho/\lambda$, where $\rho=\dot \varphi^2/2 +V(\varphi)$ is
the energy density of a scalar field (so as $\rho_b=\rho+\lambda$).
We will study two limiting cases $\rho/\lambda \ll 1$
and $\rho/\lambda \gg 1$ separately.

In the former case (a low-energy regime) expanding $(\rho+\lambda)^2$ and
neglecting $\rho^2$ term in comparison with
$\rho \lambda$, we get the standard linear dependence between Hubble parameter
square and the matter density \cite{Csaki,Cline}.
Introducing an effective 4-dimensional Planck mass $m_{P}^2=48 \pi/(k^4
\lambda)$,
the equation (8) can be rewritten in a form analogous to (3) with
the 4-dimensional Planck mass and rescaled $\tilde C = 18/(k^4
\lambda) C$:

\begin{equation}
\frac{3 m_P^2}{8 \pi}(\frac{\dot a^2}{a^2} - \frac{\tilde C}{a^4}) = \rho,
\end{equation}

It is clear that the second term in the LHS resembles the spatial
curvature in the
case of $C<0$, however, with different power-law dependence on $a$. The
question we
should answer is whether this difference is crucial for  existence of
the transient chaos in this system.

It is rather easy to show that the possibility of a bounce does not
depend significantly
on the particular form of a "curvature-like" term $C/a^{p}$ in the LHS
of eq. (9)
for an arbitrary positive $p$. Indeed,
we still have a boundary $a \sim V(\varphi)^{1/p}$ (the analog of (4)),
and the analysis
similar to \cite{we} shows that the equation for the Page points has the
form
$V'/V=Const$, where the constant depends on $C$ and $p$. This
indicates that bounces are possible for any power-law potentials (if
$\varphi$
is large enough) and can disappear for exponentially steep potentials,
as in the
closed Universe described by eqs (1)-(3). Thus, when we change $p$ in a
generalization
of the curvature term, bounce properties of the model remain
qualitatively unchanged.

However, the second condition for the chaotic dynamics -- transitions
from expansion
to contraction -- appears to be sensitive to the power index $p$. It is
clear from (9) that
a transition to contraction never happens if the
matter
density $\rho$ decreases less rapidly than $a^{-p}$ at the expansion stage.
It is well-known  that a late-time
regime for the scalar field with the potential $V \sim \varphi^n$
is damping oscillations with the effective equation of state in the form
$p=\frac{n-2}{n+2} \rho$ \cite {turner}. It means, in particular, that a
massive scalar field
($V=m^2 \varphi^2/2$) behaves like dust at the oscillatory stage ($\rho
\sim a^{-3})$,
while a self-interacting scalar field ($V= \lambda \varphi^4$) has the
equation of state
of an ultra-relativistic fluid ($\rho \sim a^{-4}$). As the dark
radiation in the RS brane
cosmology decreases as $a^{-4}$, we immediately see that oscillations of a
massive scalar field
can not be followed by the contraction epoch, and this brane Universe
will expand forever.

Numerical integrations of the system (7)-(8) for the massive scalar
field in the low-energy regime
indicates the absence of chaos. A trajectory starting from the point of
maximal expansion can be of two clearly distinguished types:

\begin{itemize}
\item A trajectory directly falling into a singularity
\item A trajectory which has a bounce and after that reaches $a \to
\infty$ regime.
\end{itemize}
Trajectories with a point of maximal expansion {\it after} bounce have not
been found.
The boundary of basins in the initial condition space leading to
this two different possibilities (singularity or eternal expansion)
is sharp without any fractal structure. This means that the
dynamics is regular (more about this method see, for example, in
\cite{fractals}).

In the case of a self-interacting scalar field its energy remains
proportional to the "dark
radiation", so a late-time recollaps of the brane Universe remains
impossible. The numerical
results for the $V=\lambda \varphi^4$ potential are qualitatively the
same as for the massive
scalar field. Only for potential $V \sim \varphi^n$ with $n \ge 6$ (the
potential
$V \sim \varphi^6$ corresponds to asymptotic equation of state in the
form $p=\rho/2$,
and leads to the energy density proportional to $a^{-4.5}$) a recollaps of
a flat brane Universe becomes inevitable, and we get the same picture as
for a closed Universe
without "dark energy".

We conclude that in the low-energy brane regime with a negative "dark
radiation" the
transient chaos is absent for a massive and self-interacting scalar
fields, and only
for power-law potentials with the index $n \ge 6$ we have a chaotic
regime, similar to
the positive spatial curvature case.

In the high-energy regime the equations of motion are

\begin{equation}
\frac{\ddot a}{a}+\frac{\dot a^2}{a^2}=-\frac{k^4}{36}
\left(\frac{\dot \varphi^4}{2}+\dot \varphi^2 V\right)
\end{equation}

\begin{equation}
\frac{\dot a^2}{a^2}-\frac{C}{a^4}=\frac{k^4}{36}\left(\frac{\dot
\varphi^2}{2}
+V\right)^2
\end{equation}

The matter part of the RHS of equations (11) is different from (9),
while the "dark
radiation" term $C/a^4$ remains unchanged. This leads to a situation,
qualitatively different
from the regime described above. Now even in the case of massive scalar field the
first item
in the RHS of (11) falls more rapidly than the "dark radiation",
providing an ultimate
recollaps. Our numerical
results for the potential $V=m^2 \varphi^2/2$ 
confirm existence of a transient chaos. Moreover,
we noticed
that for sufficiently large negative $C$ the structure of trajectories
becomes similar
to the structure described for a shallow scalar field potentials in the
standard positive
spatial curvature case. In \cite{shallow} we denote this situation as a
"strong chaos" regime,
however in the absence of unambiguous measure of chaos it is better to
call it " less regular
chaos". It's structure requires further studies.

For steeper potential the energy density during scalar field
oscillations fall even more
rapidly, and the conditions for a chaos are satisfied as well. The only
difference is that
the resulting chaos is of "classical type" (we have not found the "less
regular" chaos for any
power-law potential steeper than the quadratic one).

We have studied transient chaos on a flat isotropic brane with a scalar
field and
a negative "dark radiation" term. Our results for power-law scalar field
potentials
$V(\varphi) \sim \varphi^n$ can be summarized as follows:
\begin{itemize}
\item Low-energy regime. No chaos for $n \le 4$, classical transient
chaos for $n > 4$.
\item High-energy regime. "Less regular chaos" for $n=2$, classical
transient chaos
for $n > 2$.
\end{itemize}
On the other hand, the upper bound for possible steepness of the
potential remains the same
as for the closed brane Universe.

The complete picture of a transient chaos in brane cosmology with a
scalar field is
more complicated in comparison with these two limiting cases. In
particular, both
future outcomes
(eternal expansion and a new point of maximal expansion) are possible
after bounce,
depending on initial conditions and brane tension $\lambda$. We leave
this the most
general case to a future work.

\section*{Acknowledgments}

This work is supported by RFBR grant 05-02-17450 and
scientific school grant 2338.2003.2 of the Russian Ministry
of Science and Technology.


\begin{thebibliography}{99}
\bibitem{Page} Page D.N., Class. Quant. Grav. {\bf 1}, 417 (1984).
\bibitem{tran} Kantz H. and Grassberger P., Physica D {\bf 17}, 75 (1985).
\bibitem{Gaspard} Gaspard P. and Rice S.A., J. Chem. Phys. 
{\bf 90}, 2225 (1989).
\bibitem{Star} Starobinsky A.,  Sov. Astron. Lett. {\bf 4}, 82 (1978).
\bibitem{Cornish} Cornish N.J. and Shellard E.P.S., Phys. Rev. Lett.
{\bf 81}, 3571 (1998).
\bibitem{four} Kamenshchik A., Khalatnikov I., Savchenko S. and Toporensky A.,
Phys. Rev. {\bf D59}, 123516 (1999).
\bibitem{shallow} Pavluchenko S. and Toporensky A.,  Gravitation and
Cosmology {\bf 6}, 241 (2000).
\bibitem{Easther} Easther R. and Maeda K., Class. Quant. Grav. 
{\bf 16}, 1637 (1999).
\bibitem{Levin} Barrow J. and Levin J., Phys. Rev. Lett. {\bf 80},
656 (1998).
\bibitem{Galtsov} Dyadichev V., Galtsov D. and Moniz P.,  Phys. Rev.
{\bf D72}, 084021 (2005).
\bibitem{conformal} Calzetta E. and El Hasi C.,  Class. Quant. Grav.
{\bf 10}, 1825 (1993).
\bibitem{conformal2} Bombelli L., Lombardo F. and Castagnino M.A., J. Math. Phys. 
{\bf 39}, 6040 (1998).
\bibitem{eti} Castagnino M.A., Giacomini H. and Lara L.,  Phys.
Rev. {\bf D63}, 044003 (2001).
\bibitem{Motter} Motter A. and Letelier P. S.,  Phys. Rev. {\bf D65}, 068502 (2002).
\bibitem{we} Toporensky A., Int. J. Mod. Phys. {\bf D8}, 739 (1999).
\bibitem{KKT} Kamenshchik A., Khalatnikov. I. and Toporensky A., 
Int. J. Mod. Phys. {\bf D6}, 673 (1997).
\bibitem{recent} Toporensky A., SIGMA {\bf 2}, 037 (2006).
\bibitem{Wheeler} Krauss L. and Turner M., Gen. Rel. Grav.
{\bf 31}, 1453 (1999).
\bibitem{Caldwell} Caldwell R., Phys. Lett. {\bf B545}, 23 (2002).
\bibitem{Szydlo} Szydlowsli M., Krawiec A. and Czaja W.,
Phys. Rev. {\bf E72}, 036221 (2005).
\bibitem{Szydlo2} Szydlowsli M., Hrycyna O. and Krawiec A.,
"Phantom cosmology as a scattering process", hep-th/0608219.
\bibitem{Langlois} Binetruy P., Deffayet C. and Langlois D., Nucl.
Phys. 2000, {\bf B565}, 269 (2000).
\bibitem{Langlois2} Binetruy P., Deffayet C., Ellwanger U. and Langlois
D.,  Phys. Lett. {\bf B477}, 285 (2000).
\bibitem{Csaki} Csaki C., Graesser M., Kolda C. and Terning J., 
Phys. Lett. {\bf B462}, 34 (1999).
\bibitem{Cline} Cline J., Grossjean C. and Servant G., Phys. Rev.
Lett. {\bf 83}, 4245 (1999).
\bibitem{turner} Turner M., Phys. Rev. {\bf D28}, 1243 (1983).
\bibitem{fractals} Cornish N. and Levin J., Phys. Rev. {\bf D53}, 
3022 (1996).
\end{thebibliography}
\end{document}